\documentclass[conference]{IEEEtran}
\usepackage[cmex10]{amsmath}
\usepackage{nomencl}
\usepackage{amsmath}
\usepackage{comment}
\usepackage[table]{xcolor}
\usepackage{tikz}
\usepackage{gensymb}
\usepackage{bm}
\usepackage{mathtools}
\usepackage{soul}
\usepackage{multirow}
\usepackage{booktabs,colortbl}
\usepackage{umoline}
\usepackage[utf8]{inputenc}
\usepackage[T1]{fontenc}
\usepackage{ wasysym }
\usepackage{balance}
\usepackage{hhline}
\usepackage{graphicx}
\usepackage{array}
\usepackage{subfigure}
\usepackage{verbatim}
\usepackage{cleveref}
\usepackage{color}
\usepackage{relsize}
\usepackage{tikz}
\usepackage{cite}
\usepackage{comment}
\usepackage{epstopdf}
\usepackage{fancyhdr}
\usepackage{graphics}
\usepackage{float}
\usepackage{tablefootnote}
\usepackage{tabularx}
\usepackage{threeparttable}
\usepackage{algorithm,algorithmic}
\usepackage{dblfloatfix}

\renewcommand{\baselinestretch}{0.94} 
\graphicspath{{./figures/}}

\ifCLASSOPTIONcompsoc

\else

\fi

\ifCLASSINFOpdf

\else

\fi

\newcommand{\hlblue}[1]{{\color{black} #1}}

\begin{document}
	%
	\title{Real-time Modeling of Offshore Wind Turbines for Transient Simulation and Studies}
			
	%
	%

\makeatletter
\newcommand{\linebreakand}{%
\end{@IEEEauthorhalign}
\hfill\mbox{}\par
\mbox{}\hfill\begin{@IEEEauthorhalign}
}
\makeatother

	\author{
		\IEEEauthorblockN{Thai-Thanh~Nguyen}
		\IEEEauthorblockA{	\textit{ECE Department} \\
										\textit{Clarkson University}\\
										Potsdam, NY, USA \\
										tnguyen@clarkson.edu}
		\and
		\IEEEauthorblockN{Tuyen~Vu}
		\IEEEauthorblockA{	\textit{ECE Department} \\
										\textit{Clarkson University}\\
										Potsdam, NY, USA \\
										tvu@clarkson.edu}
		\and	
		\IEEEauthorblockN{Thomas~Ortmeyer}
		\IEEEauthorblockA{	\textit{ECE Department} \\
										\textit{Clarkson University}\\
										Potsdam, NY, USA \\
										tortmeye@clarkson.edu} 
		\linebreakand
		
		\IEEEauthorblockN{George~Stefopoulos}
		\IEEEauthorblockA{	\textit{Advanced Grid Innovation Lab for Energy} \\
										\textit{New York Power Authority}\\
										White Plains, NY, USA \\
										Georgios.Stefopoulos@nypa.gov}
		\and
		\IEEEauthorblockN{Greg~Pedrick}
		\IEEEauthorblockA{	\textit{New York Power Authority}\\
										White Plains, NY, USA \\
										Gregory.Pedrick@nypa.gov}
		\and
		\IEEEauthorblockN{Jason~MacDowell}
		\IEEEauthorblockA{	\textit{GE Power}\\
										Schenectady, NY, USA \\
										jason.macdowell@ge.com}
	}

	\markboth{IEEE Trans. Sustainable Energy}%
	{Shell \MakeLowercase{\textit{et al.}}: Bare Demo of IEEEtran.cls for IEEE Journals}

	\maketitle
	
	\begin{abstract}

		The real-time models of offshore wind turbines are proposed in this study, which are developed in compliance with the Western Electricity Coordinating Council (WECC) standard to meet the industry requirements. In addition to basic functionalities of the generic WECC turbine model such as power curtailment and voltage ride through, the sequence current control and sequence current limit are designed for future electromagnetic transient (EMT) testing and control of offshore wind farms. Both average-value and switching detailed models are modeled in the Opal-RT simulator. Both balanced and unbalanced faults are studied to show the feasibility of the proposed turbine models. The models are validated against the WECC second-generation generic wind turbine model. The active and reactive power results for low-voltage ride through cases validated the turbine model's performance against WECC generic model in the balanced system. In addition, the models provide extended capability in mitigating the active power oscillation during unbalanced fault conditions.

	\end{abstract}
	
	
	%
	\IEEEpeerreviewmaketitle
	
	
	\section{Introduction}
	
	\IEEEPARstart{O}{ffshore} wind energies have recently attracted considerable attention because of the necessity for decarbonisation and the decrease of fossil fuels, as well as their ability to produce more wind power at a higher efficiency than onshore wind energy systems \cite{IRENA}. Thanks to the fast advancement of offshore wind turbine technology, the offshore wind system is now cost-competitive with direct-drive multi-megawatt turbines. The problem of gearbox failure is addressed by advanced technology of the direct-drive turbine systems, resulting in an increase in the usage of high-power direct-drive turbines capable of producing up to 15 MW \cite{gaertner2020definition, Vestas15MW}. Offshore wind projects are getting bigger as developer experience and industry maturity significantly increase. Modeling of offshore wind turbines that meets industry standards is critical for assessing the performance of offshore wind projects.	
		
	Various detailed electromagnetic transient (EMT) models of wind turbines have been presented, which have been developed using offline simulation tools like  PSCAD/EMTDC, EMTP, and MATLAB/Simulink. These models, while allowing for the modeling and simulation of complicated turbine models for transient simulation, are not real-time and are generally sluggish. A generic EMT-type turbine model with a 2~MW capability has been presented in \cite{8396277}, with the proposed model being validated against field testing of the ENERCON E-82 2.3~MW turbine. In \cite{5262959}, a 5~MW PMSG wind turbine model with low-voltage ride-through capabilities has been introduced, which is assessed using the US grid code. However, the proposed models in \cite{8396277} and \cite{5262959} were designed based on the phasor domain (positive sequence),which may result in improper protection system functioning under some imbalanced situations owing to the lack of negative sequence current contribution during the fault. Furthermore, during unbalanced fault conditions, the presence of negative sequence voltage produces second-order harmonic oscillations in output active power and DC-link voltage. \cite{6520231}. A 1.5~MW wind turbine model based on symmetrical components was presented in \cite{Karaagac2019EMT}, with the capacity to inject negative sequence current during unbalanced faults. The sequence-based detailed switching (DSW) and averaged-value (AVG) models were also examined, but neither was implemented in real time. In addition, the AVG model could not adequately capture the DSW model's dynamic responses.
	According to the CPU-usage measurements in \cite{Karaagac2019EMT}, the DSW model used 144.7~s of CPU time for a 1~s simulation time period, whereas the AVG model consumed 28.8~s. Therefore, there is still a research need to fill, as well as a requirement to develop and test real-time turbine models.
	
	Real-time models of wind turbines have been presented in the literature \cite{9067370,6869686,5342521,5345723}. These models, however, depict small wind turbines with capacities of 1.5~MW and 2~MW, which are insufficient for offshore wind farms that incorporate turbines power of more than 6~MW \hlblue{due to the lack of voltage-ride through capability. In addition, these models do not have the capacity to regulate negative sequences and have not been validated against any international standards.} To overcome these limitations, this study proposes real-time DSW and AVG turbine models of high-power direct-drive PMSG wind turbines for offshore wind studies. The proposed models are validated against the Western Electricity Coordinating Council (WECC) standard to ensure that they are applicable for practical studies. The main contributions of this paper are as follows:
	\begin{itemize}
		\item Real-time models of high-power direct-drive wind turbines that meet international standards are proposed. 
		\item Both detailed switching and average-value models are developed and evaluated by the real-time simulator.
		\item \hlblue{The proposed sequence-based real-time turbine models are capable of injecting negative sequence current during unbalanced faults.}
	\end{itemize}
	
	The rest of this paper is organized as follows. Detailed wind turbine model and its control system in accordance with the WECC standard are presented in Section \ref{sec:WTG_model}. Section \ref{sec:MValidation} describes the model validation against the WECC standard. Section \ref{sec:Evaluation} presents the evaluation of both DSW and AVG models under normal and abnormal conditions including balanced and unbalanced dynamic studies. Finally, the main findings of this paper is summarized in Section \ref{sec:conclusion}.
	

	\section{Wind Turbine Modeling}
	\label{sec:WTG_model}
	\subsection{Configuration of Direct-drive PMSG Wind Turbine}
	\begin{figure}[b]
		\centering
		\includegraphics[width=0.95\linewidth]{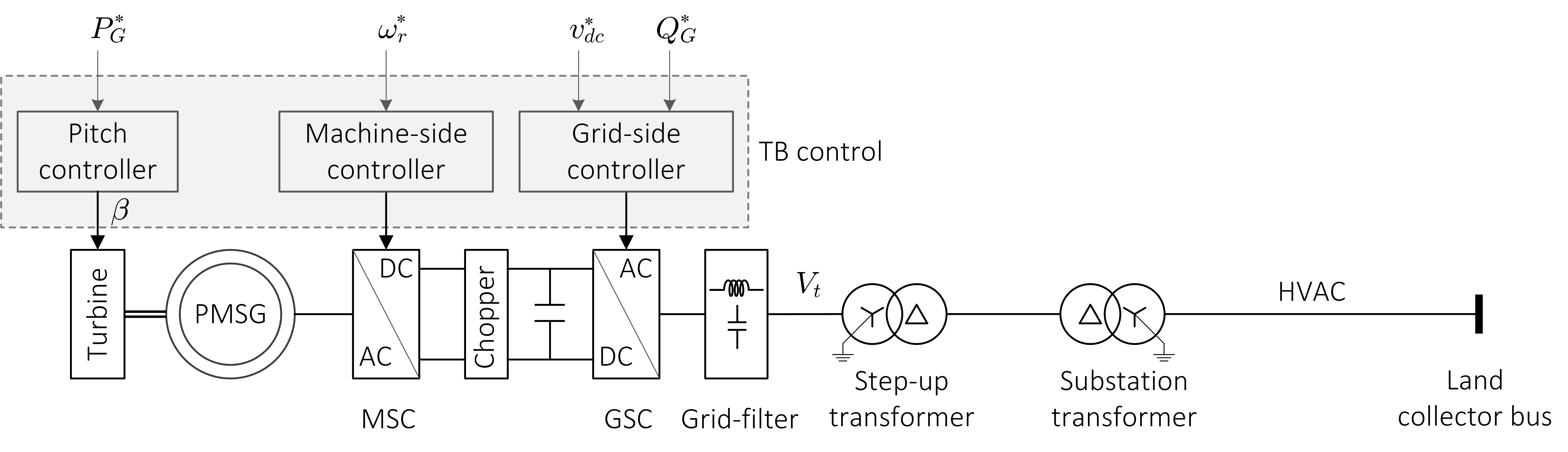}
		\caption{Direct-drive PMSG with control functions.}
		\label{fig:TB_control_sys}
	\end{figure}
	The typical configuration of the direct-drive PMSG wind turbines is depicted in Fig.~\ref{fig:TB_control_sys}, in which the wind turbine is directly connected with the rotor of the synchronous generator. The back-to-back (BTB) converter, which consists of a machine-side converter (MSC), a grid-side converter (GSC), and a DC chopper, is linked to the stator winding of the PMSG. The MSC is in charge of PMSG torque management, while the GSC is in charge of the grid's DC-link voltage and reactive power exchange. The DC chopper is connected in parallel with the DC link to maintain a stable DC-link voltage during transient conditions. There are three main controllers of the direct-drive PMSG wind turbine, which are pitch angle control, machine-side control, and grid-side control. 	
	
	\subsection{Pitch Angle Controller}
	
	The pitch angle control, as shown in Fig.~\ref{fig:pitchcontrol} is used to adjust the aerodynamic power captured from wind energy. To produce the pitch angle reference in region 1.5 ($\beta_{R1}$), a lookup table using wind speed ($v_w$) as an input is utilized. When the wind speed ($v_w$) is higher than the cut-in speed ($v_{cin}$) but lower than the rated wind speed ($v_{rated}$), the pitch angle reference ($\beta^*$) is equal to zero. The pitch angle is adjusted to manage the turbine output power when the wind speed exceeds the rated wind speed or the turbine power reference ($P^*$) is less than the rated power ($P_{rated}$). The proportional-integral (PI) regulator in conjunction with the limits magnitude and rate-of-change is used for pitch regulation.
	
	\begin{figure}[!bt]
		\centering
		\includegraphics[width=0.785\linewidth]{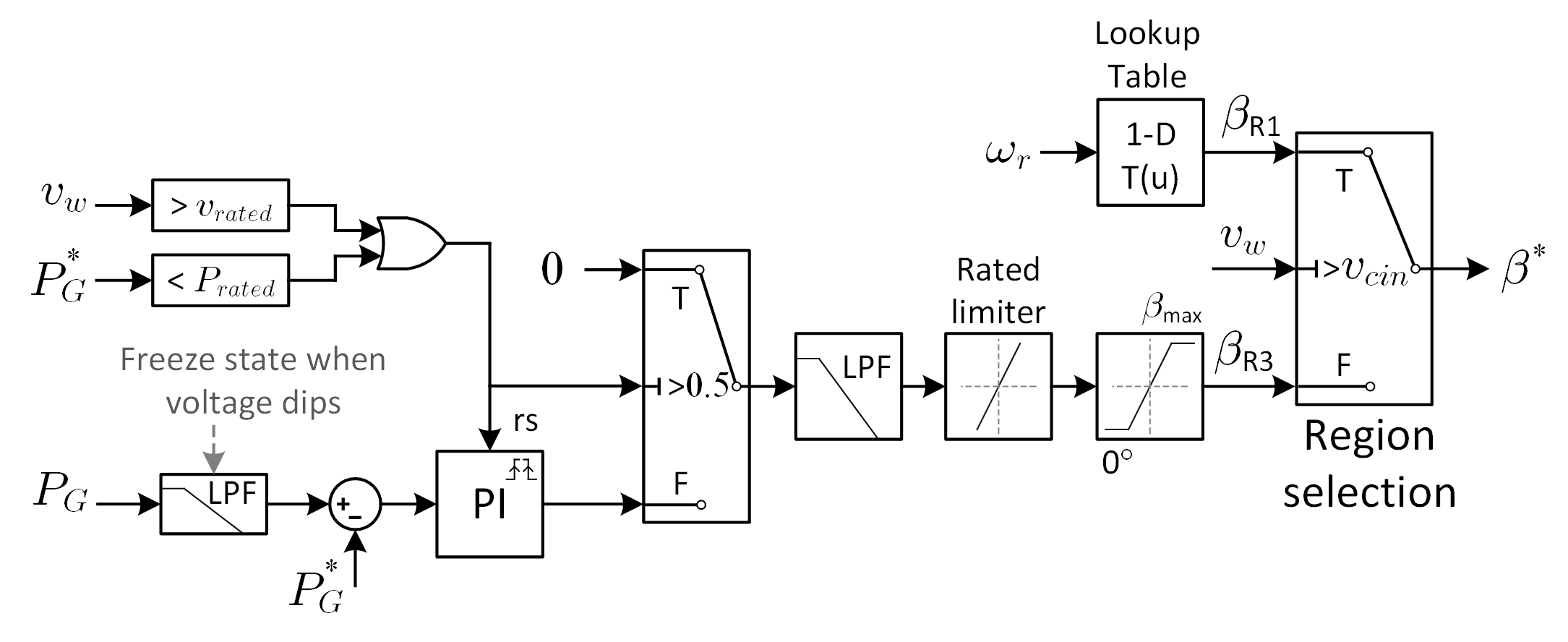}
		\caption{Schematic diagram of the pitch controller.}
		\label{fig:pitchcontrol}
	\end{figure}
	
	\subsection{Machine-side Controller}
	The MSC controller, which is shown in Fig.~\ref{fig:MSCcontrol}, includes an inner current control loop and an outer torque control loop. The current reference ($i_{Md}^*$) for the inner current loop is generated by the torque controller. To compensate for torque errors, the PI regulator with the rated limiter is employed. Depending on the wind speed, the torque reference is given by the speed control or optimal torque control. When the wind speed is less than the rated speed, the wind turbine operates in region 2, which uses the optimal torque in (\ref{eq:Topt}) to maximize the wind power. The wind turbine operates in region 3 when the wind speed is higher than the rated wind speed. The generator torque is managed in this region by a rotor speed control, which keeps the rotor speed constant at 1 pu. When the wind speed exceeds the cut-out speed in region 4, the wind turbine is turned off.
	\begin{figure}[!bt]
		\centering
		\includegraphics[width=0.785\linewidth]{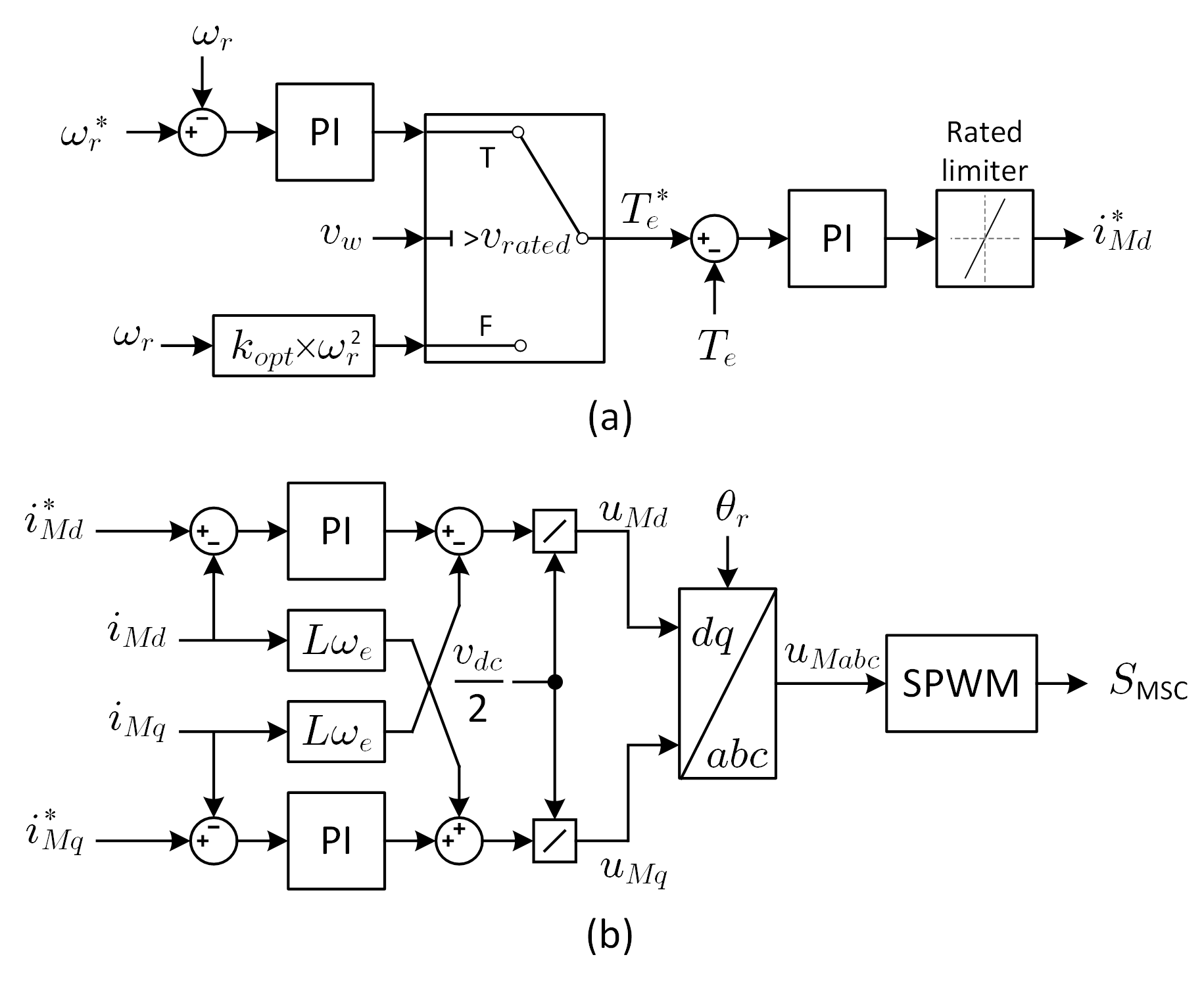}
		\caption{Control diagram of the machine-side converter: (a) Torque control loop; (b) Inner current control loop.}
		\label{fig:MSCcontrol}
	\end{figure}
	
	\begin{align}
		\label{eq:Topt}
		T_{opt} & = k_{opt}\omega_r^2, \\
		k_{opt} & = 0.5 \rho \pi R^2 C_{p}^{max} (R / \lambda_{opt})^3,
	\end{align}
	where $\lambda_{opt}$ is the optimal tip-speed ratio.
	
	\subsection{Grid-side Controller} 
	
	\begin{figure*}[h]
		\centering
		\includegraphics[width=0.75\textwidth]{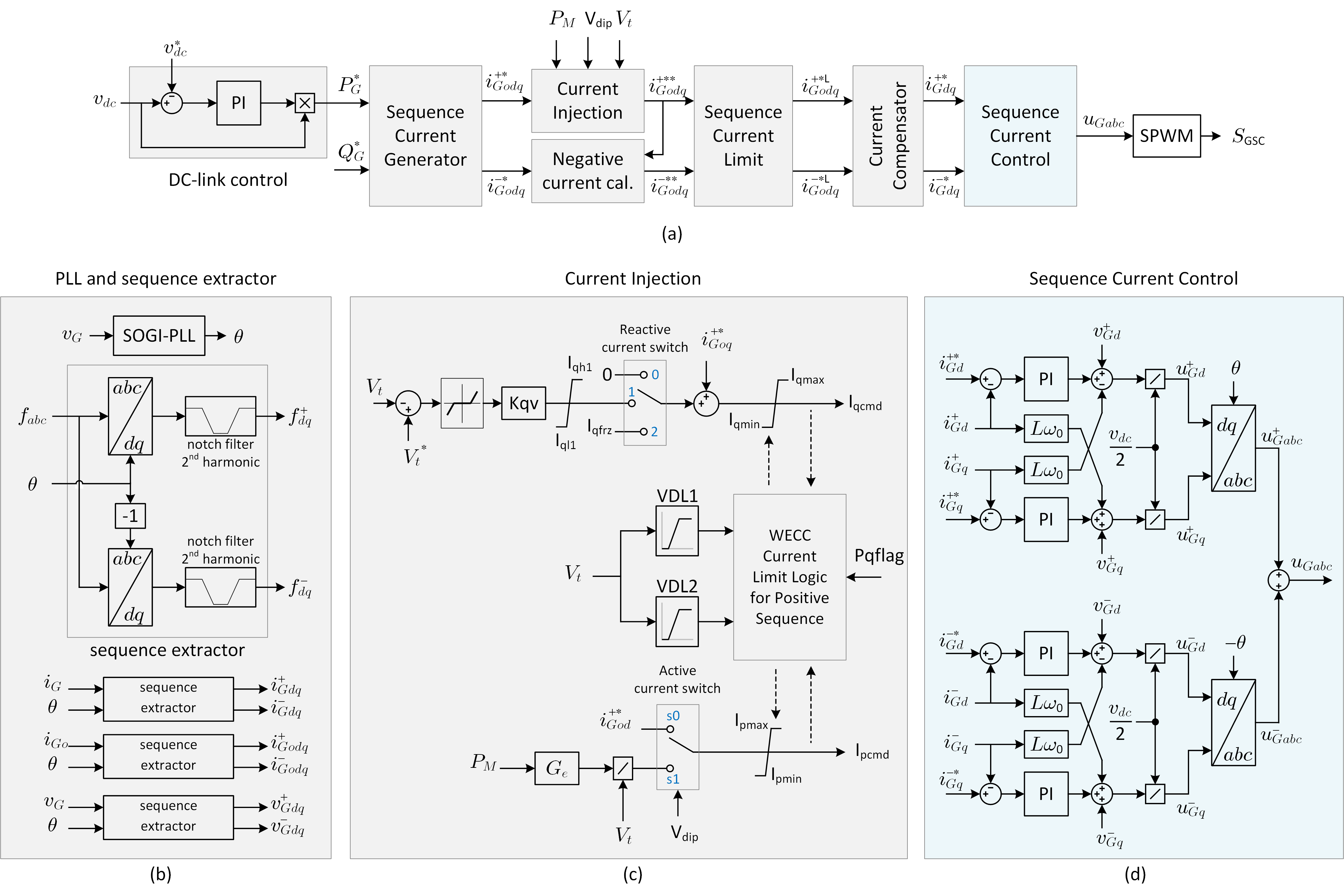}
		\caption{Control diagram of the grid-side converter: (a) Overal control diagram; (b) SOGI-PLL and sequence current and voltage extractor; (c) Active and reactive current injection; (d) Sequence current control loop.}
		\label{fig:GSC_control}
	\end{figure*}
	
	The symmetrical components of turbine voltage and current are used to design the grid-side controller for the wind turbine in this study. The positive and negative sequence components of the unbalanced three-phase voltage are converted into the synchronous rotating frame, as given by (\ref{eq:seqdq}), in which $v_{Gdq}^+$ is the positive $dq$ component while $v_{Gdq}^-$ is the negative $dq$ component \cite{1022376}.
	
	\begin{align}
		\label{eq:seqdq}
		v_{Gdqs} & = v_{Gdq}^+ e^{j\omega t} + v_{Gdq}^- e^{-j\omega t} \nonumber \\ 
		& = {2 \over 3}(v_{Ga} + av_{Gb} + a^2 v_{Gc}), \\
		\label{eq:posdq}
		v_{Gdq}^+ & = {2 \over 3}(v_{Ga}^+ + av_{Gb}^+ + a^2v_{Gc}^+) e^{-j\omega t} = v_{Gd}^+ + jv_{Gq}^+, \\
		\label{eq:negdq}
		v_{Gdq}^- & = {2 \over 3}(v_{Ga}^- + av_{Gb}^- + a^2v_{Gc}^-) e^{j\omega t} = v_{Gd}^- + jv_{Gq}^-,
	\end{align}
	where $a = e^{j{2\pi \over 3}}$; $v_G$ is the terminal voltage of converter. \\
	
	In the same way, the positive and negative sequence currents, $i_{Godq}^+$ and $i_{Godq}^-$, are defined. The sequence components of voltage and current, as provided by (\ref{eq:PQseq}), are used to compute complex power, which includes second-order harmonic oscillations of power.  

	\begin{align}
		\label{eq:PQseq}
		S & = p(t) + jq(t) = v_{Gdqs} i_{Godqs}^*, \\
		p(t) & = P + P_{c2}\cos(2 \omega t) + P_{s2}\sin(2 \omega t), \\
		q(t) & = Q + Q_{c2}\cos(2 \omega t) + Q_{s2}\sin(2 \omega t).
	\end{align}

	The control design does not take into account second-order fluctuations of reactive power \cite{8579187}. Active power oscillations, on the other hand, cause the DC-link voltage to oscillate at twice the nominal frequency. The oscillatory terms of active power ($P_{c2}$ and $P_{s2}$) are adjusted to zero to mitigate such oscillations in DC-link voltage.
	
	\begin{align}
		\label{eq:idqpnref}
		\begin{bmatrix}
			{i_{God}^{+*} } \\ {i_{Goq}^{+*} } \\ {i_{God}^{-*} } \\ {i_{Goq}^{-*} }
		\end{bmatrix}
		= {2 \over 3}
		\begin{bmatrix}
			{v_{Gd}^{+} } & {v_{Gq}^{+} } & {v_{Gd}^{-} } & {v_{Gq}^{-} } \cr {v_{Gq}^{+} } & { - v_{Gd}^{+} } & {v_{Gq}^{-} } & { - v_{Gd}^{-} } \cr {v_{Gq}^{-} } & { - v_{Gd}^{-} } & { - v_{Gq}^{+} } & {v_{Gd}^{+} } \cr {v_{Gd}^{-} } & {v_{Gq}^{-} } & {v_{Gd}^{+} } & {v_{Gq}^{+} }
		\end{bmatrix}^{-1}
		\begin{bmatrix}
			{P} \cr {Q} \cr {P_{s2} } \cr {P_{c2} }
		\end{bmatrix}
	\end{align}
	
	The overall control scheme of the grid-side converter depicted in Fig.~\ref{fig:GSC_control}(a) has the primary functions of controlling DC-link voltage and output reactive power. The DC-link voltage controller generates the active power reference ($P^*$), whereas the upper control layer, such as the wind power plant level controller, provides the reactive power reference ($Q^*$). The sequence current generator based on (\ref{eq:idqpnref}) calculates the sequence current references for the inner sequence current control loop using the active and reactive power references as inputs. The second-order generalized integrator (SOGI) phase-locked loop (PLL), which can achieve accurate phase-locking under unbalanced conditions, is used to calculate the phase angle of terminal voltage ($\theta$), which is then used to extract the sequence components of current and voltage, as shown in Fig.~\ref{fig:GSC_control}(b). 
	
	The low-voltage ride through requirement is addressed by the current injection function that manages the active and reactive current injected into the grid under low-voltage conditions, as shown in Fig.~\ref{fig:GSC_control}(c). The present injection function is based on the WECC generic turbine model \cite{remtf2014wecc}. Two look-up tables (VDL1 and VDL2) in \cite{remtf2014wecc} determine the limits on the total active and reactive currents injected into the grid under low voltage conditions. At very low voltages, the VDL1 and VDL2 curves allow for zero active and reactive currents. In this study, the WECC current limit logic is modified for positive sequence currents. Because the GSC's active current might be negative to maintain the DC-link voltage, the updated WECC current limit logic differs from the original in terms of the lower limit of active current injection ($I_{pmin}$).
	
	The voltage dip is monitored ($v_{dip} = 1$) when the terminal voltage ($v_t$) falls below a predetermined value, and the current injection function controls the amount of injected active and reactive current into the grid under low voltage conditions. When a voltage-dip condition is recognized, the statuses of the reactive current switch and active current switch are changed from 0 to 1. The injected reactive current is proportional to the voltage drop, but the injected active power is determined by the mechanical power. In order to control the injected reactive current in relation to voltage-dip conditions, a deadband is utilized.
	
	\begin{figure}[!bt]
		\centering
		\includegraphics[width=3.5in]{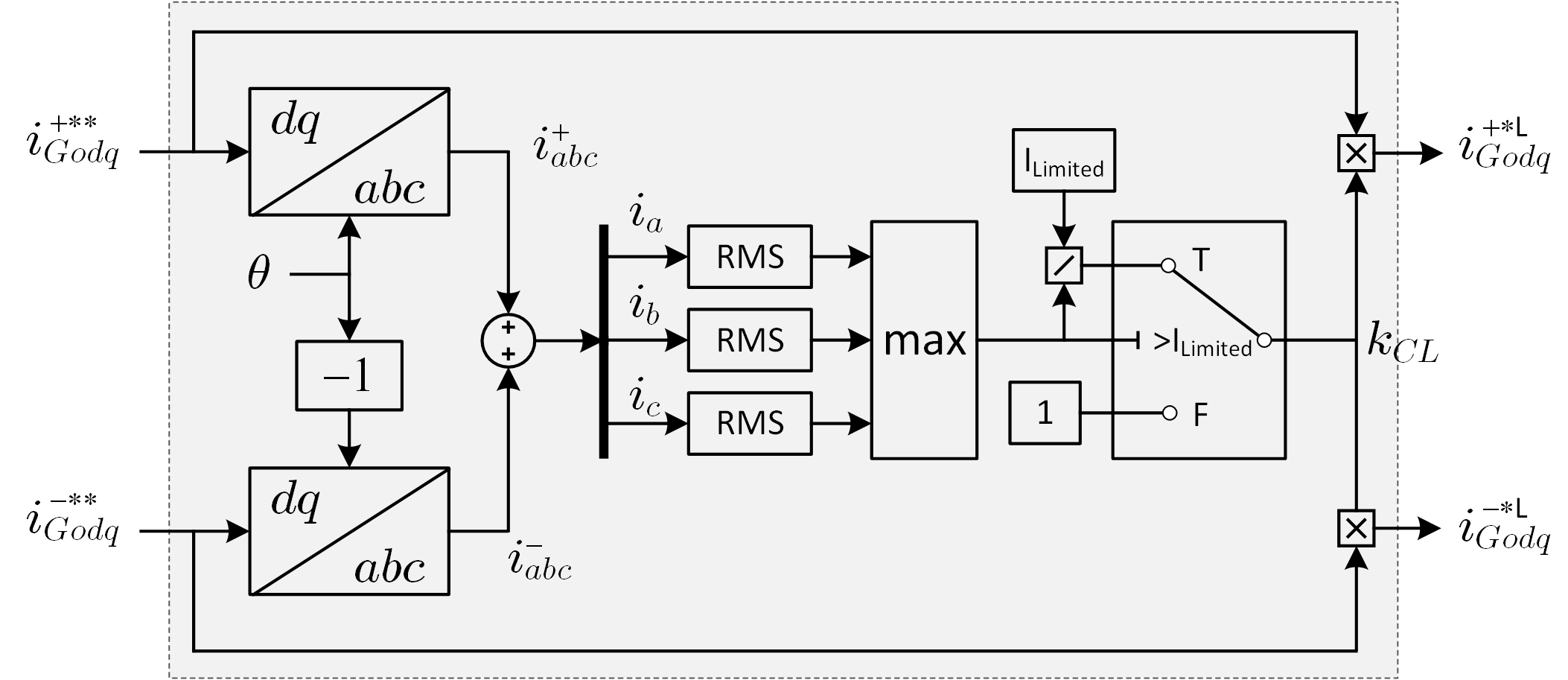}
		\caption{Schematic diagram of the sequence current limit.}
		\label{fig:sequence_limit}
	\end{figure}

	In addition, the sequence current limit function is utilized to restrict the RMS value of each phase current. Fig.~\ref{fig:sequence_limit} illustrates the schematic diagram of the sequence current limit function. The symmetrical components of current references transformed back to phasor components. Each phase's RMS value is computed accordingly. The current limit gain ($k_ {CL}$) is one during normal operation when the converter current is less than the restricted value. If the largest RMS value among the three phases exceeds the limited value during a disturbance, the current limit function starts working to lower the current references by adjusting $k_{CL}$.  
	
	The current compensator block is used to convert the inductor filter current references $i_{Gdq}^*$ from the limited positive and negative output load current references $i_{Godq}^{*L}$. The inductor filter current references are used as inputs to the sequence current control, which comprises of positive and negative sequence current controllers. The sequence current control diagram is depicted in Fig.~\ref{fig:GSC_control}(d), in which $L$ is the filter inductance, $\omega_0$ is the grid angular frequency, and $v_{dc}$ is the DC-link voltage.
	
	\section{Model Validation}
	\label{sec:MValidation}
	
	\subsection{Test System}
	
	In this paper, the detailed 15~MW model is validated against the WECC generic model to ensure that the proposed model is appropriate for the practical study of the interconnected wind farm system. The validation setup shown in Fig.~\ref{fig:WTGMVsetup} includes a controlled voltage source to mimic the voltage dip that occurs on the utility side. The measured voltage dip at the terminal of the wind turbine generator (WTG) is shown in Fig.~\ref{fig:VsMV}, for four different cases. The wind turbine models are validated in each case under various wind speed conditions. These voltage instances are utilized as validation inputs in the WECC model. To compare with the WECC model, the active and reactive power responses of detailed models are recorded for various voltage situations. Table \ref{table:WTGparameters} shows detailed parameters of the wind turbine generator.
	
	\begin{figure}[!bt]
		\centering
		\includegraphics[width=0.78\linewidth]{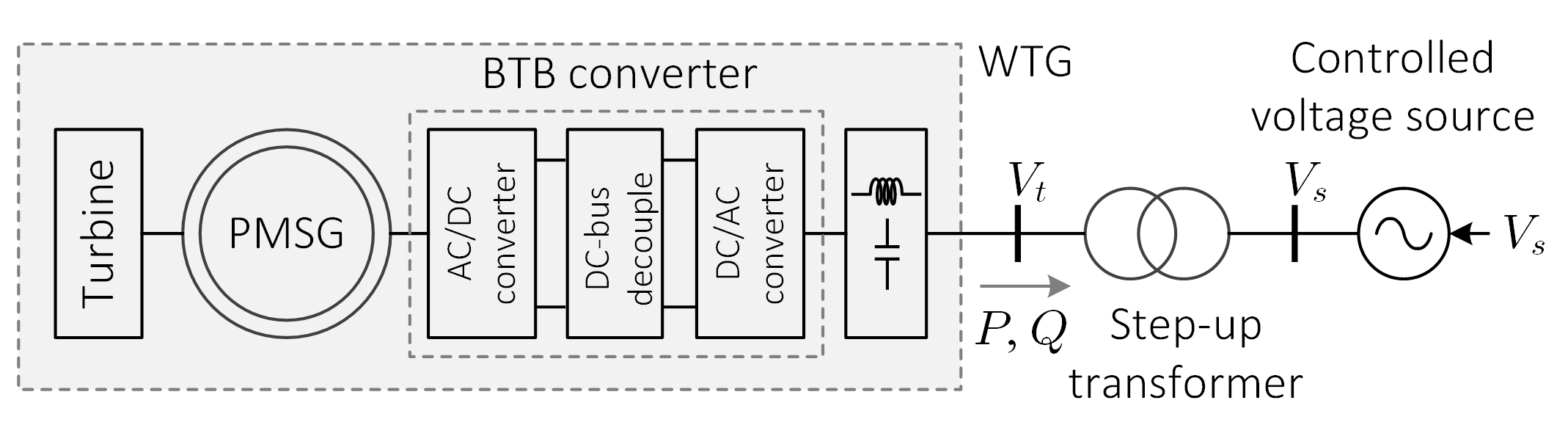}
		\caption{Setup for model validation.}
		\label{fig:WTGMVsetup}
	\end{figure}
	\begin{figure}[!bt]
		\centering
		\includegraphics[width=0.78\linewidth]{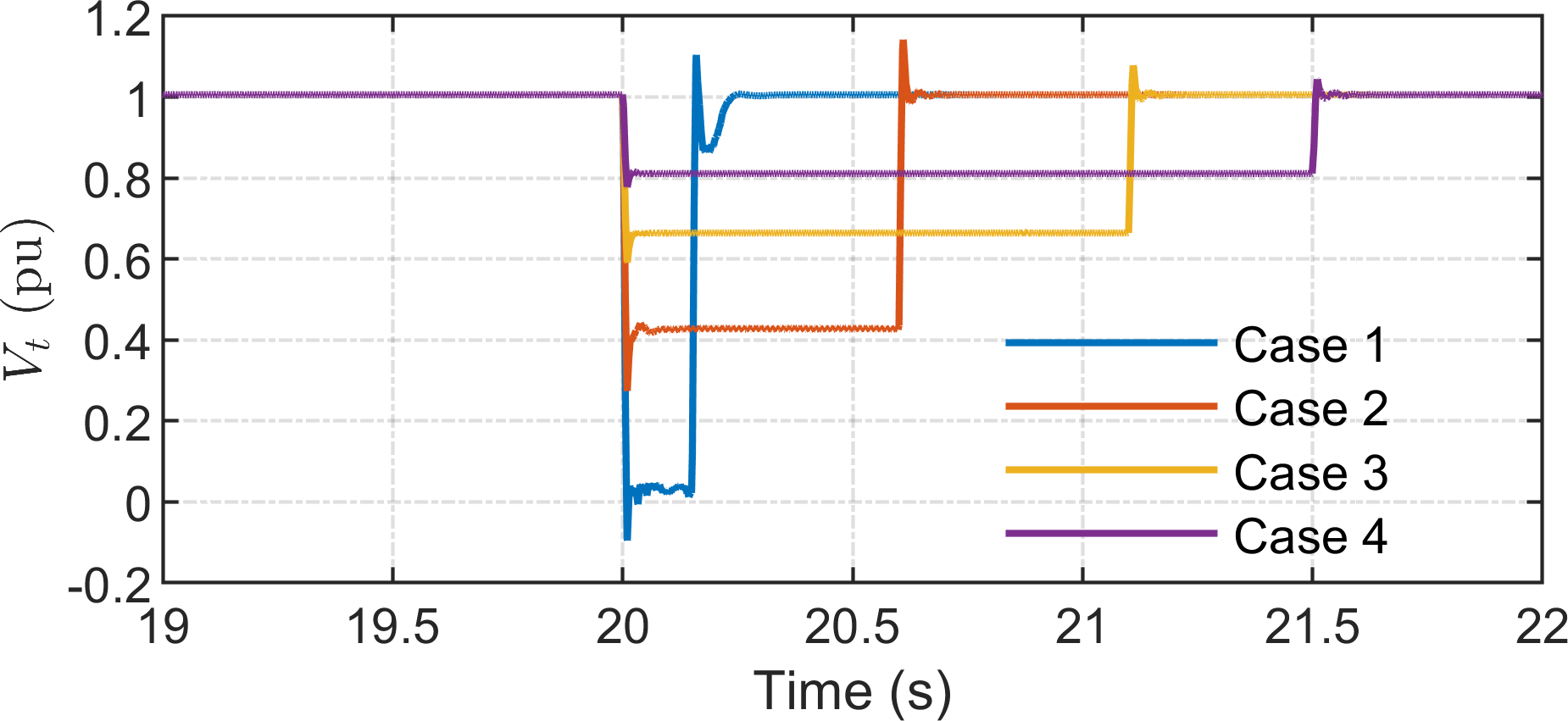}
		\caption{Voltage dip measured at the WTG terminal.}
		\label{fig:VsMV}
	\end{figure}
	\begin{table}[!bt]
		\renewcommand{\arraystretch}{1.3}
		\caption{Parameters of 15~MW wind turbine model.}
		\label{table:WTGparameters}
		\centering
		\begin{tabular}{c | c}
			\hline\hline
			Components & Parameters \\ 
			\arrayrulecolor{gray}\hline
			\multirow{3}{*}{Step-up transformer} 
			& Rated power: 18 MVA \\
			& Rated voltage (RMS line): 4 $\mid$ 66 kV \\
			& Leakage reactance: 0.1 pu \\
			
			\hline
			
			\multirow{5}{*}{BTB converter}
			& Rated power: 15 MW\\ 	 
			& DC link voltage: 10 kV\\ 	
			& DC link capacitance: 4000 $\mu$F\\	  
			& Output $LC$ filter: 0.275 mH; 1024 $\mu$F\\
			& Switching frequency: 3 kHz\\
			
			\hline
			
			\multirow{7}{*}{PMSG} 
			& Rated power, $P_{rated}$: 15 MVA\\ 	
			& Number of pole pairs, $p$: 162\\	  
			& Stator resistance, $R_s$: 0.0368 $\ohm$\\
			& Stator $d$-axis inductance, $L_d$: 0.0087 H\\
			& Stator $q$-axis inductance, $L_q$: 0.0058 H\\
			& Rated speed, $\omega_r$: 0.8 rad/s\\
			& Total reflected inertia, $J$: $3.16 \times 10^8$ kgm$^2$\\
			
			\hline
			
			\multirow{2}{*}{Wind turbine} 
			& Rated wind speed, $v_{rated}$: 12~m/s\\ 	
			& Turbine rotor diameter, $D$: 236 m\\
			
			\arrayrulecolor{black}\hline\hline
		\end{tabular}
	\end{table}
	
	\subsection{Model Validation}
	
	\begin{figure}[!bt]
		\centering
		\subfigure[Case 1]
		{
			\includegraphics[width=0.35\linewidth]{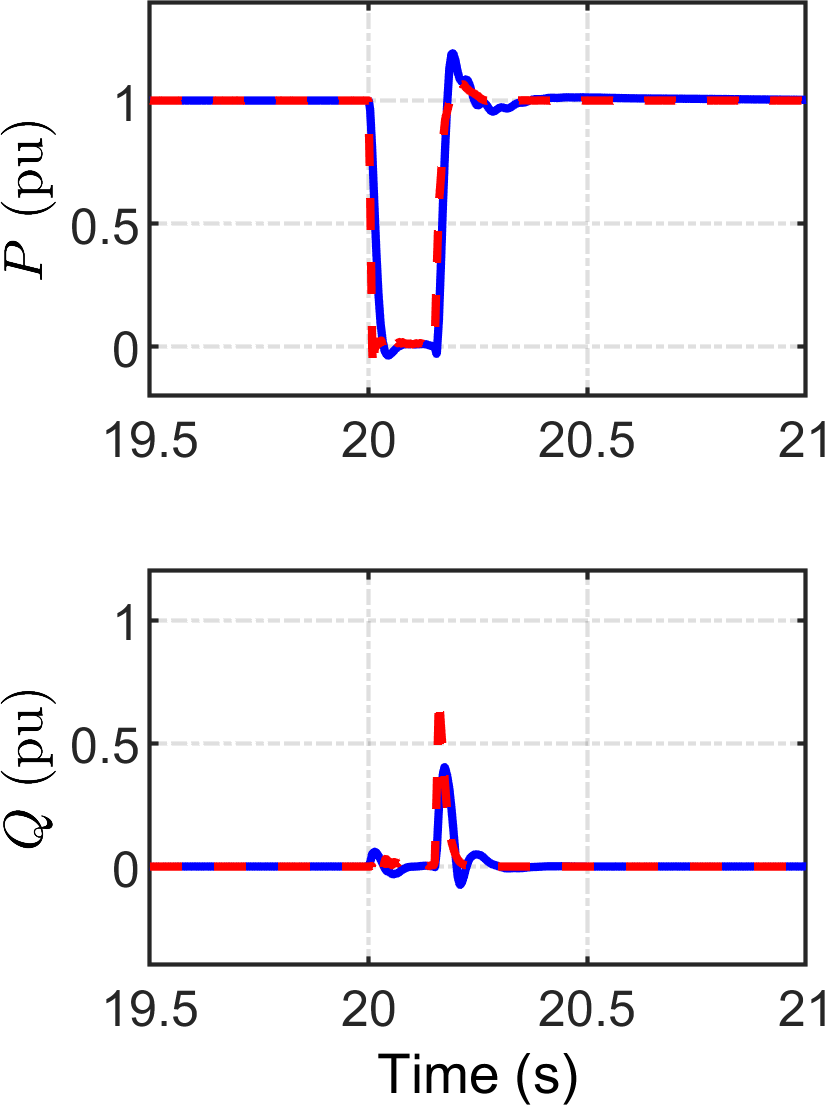}
		} \quad	
		\subfigure[Case 2]
		{
			\includegraphics[width=0.35\linewidth]{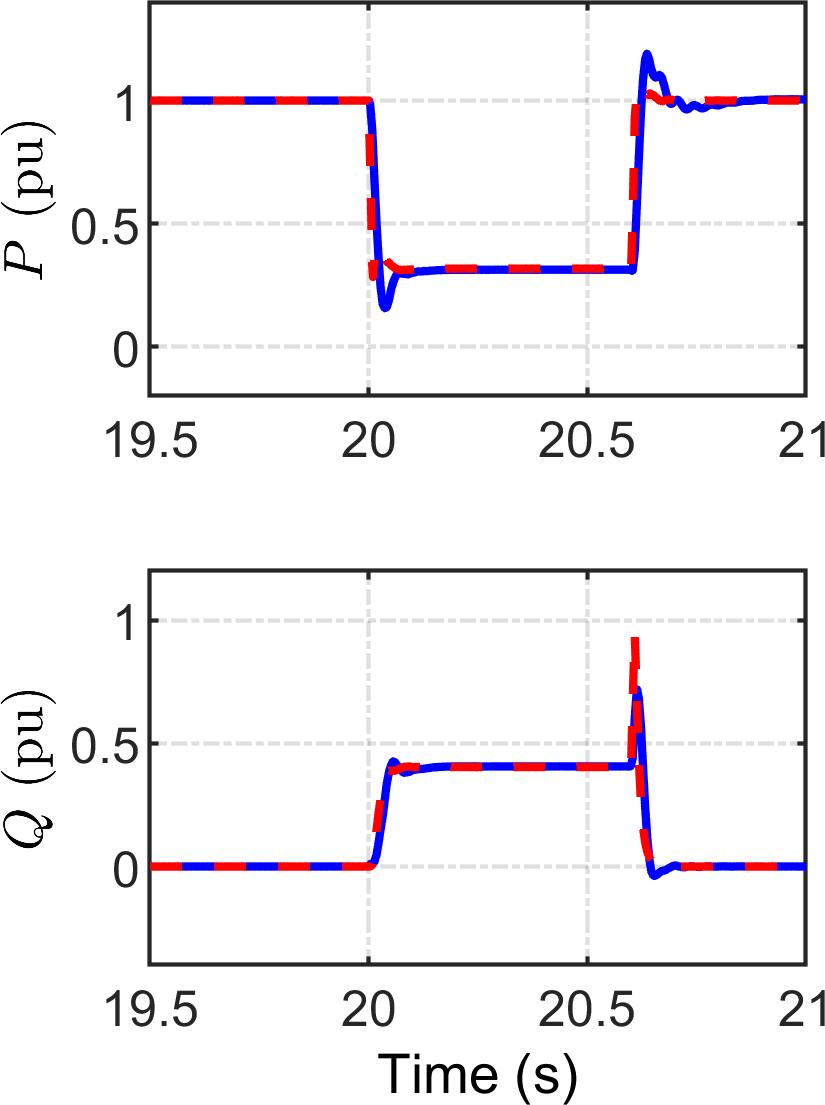}
		} 
		
		\subfigure[Case 3]
		{
			\includegraphics[width=0.35\linewidth]{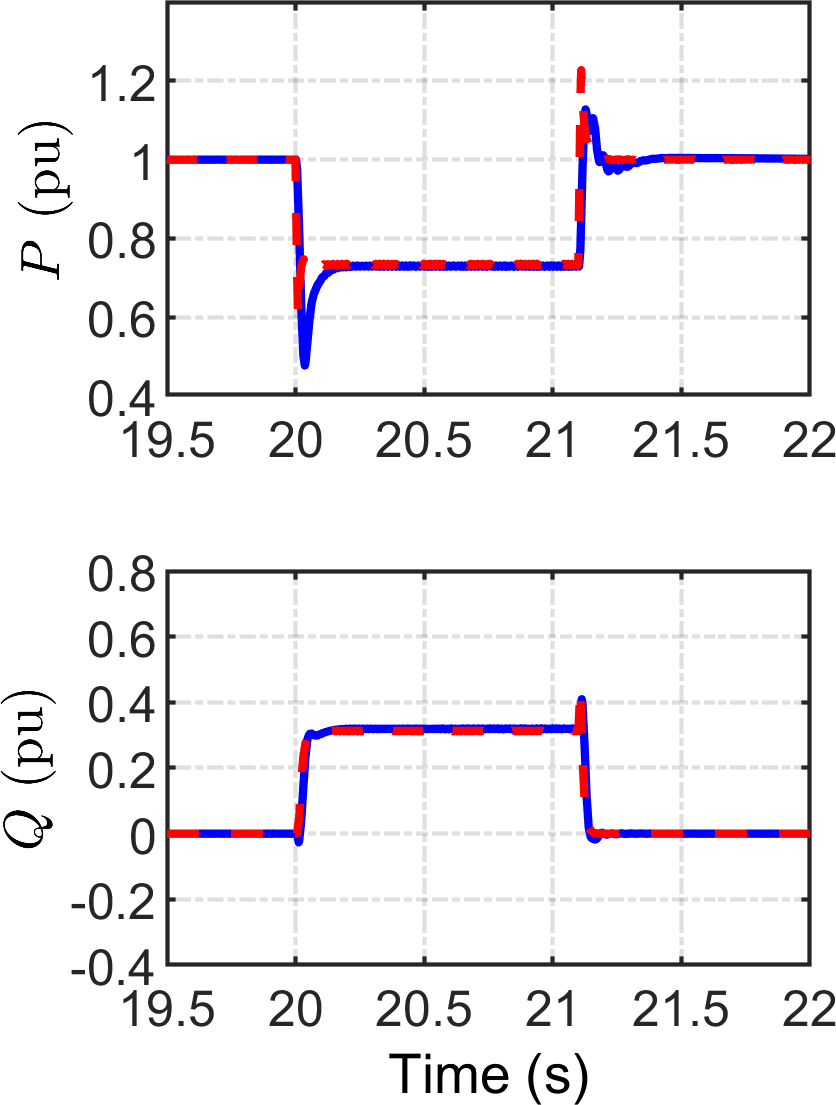}
		} \quad	
		\subfigure[Case 4]
		{
			\includegraphics[width=0.35\linewidth]{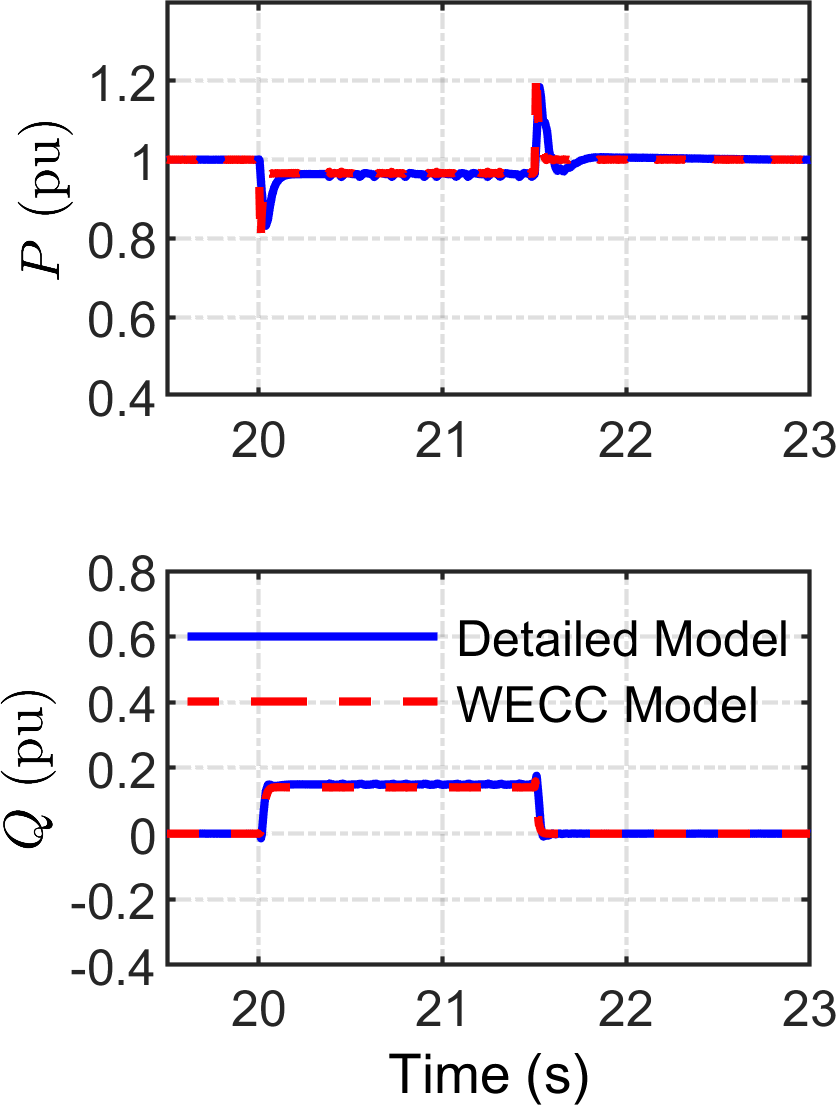}
		}
		\caption{Model validation under wind speed of 15~m/s.}
		\label{fig:MV15mps}
	\end{figure}

	The validation results for a wind turbine operating in region~3 with a wind speed of 15 m/s are shown in Fig.~\ref{fig:MV15mps}. The validation results are presented in Fig.~\ref{fig:MV15mps} for situations when the wind turbine has a rated power output. The proposed wind turbine model, like the WECC general model, has the capacity to inject electricity during voltage dips. Turbine power is forced to zero in case of significant voltage drop, as shown in Fig.~\ref{fig:MV15mps}(a). In cases 3 and 4, where the voltage drops to 0.6~pu and 0.8~pu, respectively, the active power of the turbine reduces and the reactive power rises to 0.32~pu and 0.18~pu, respectively. It can be observed that the performance of detailed model matches that of the WECC generic model.
	
	\section{Evaluation of DSW and AVG Models}
	\label{sec:Evaluation}
		
	\subsection{Balanced Fault Studies}
	The wind turbine models are assessed in the presence of a balanced fault on the high side of the step-up transformer from Fig.~\ref{fig:WTGMVsetup}. The evaluation results are depicted in Figs.~\ref{fig:VIabc_DSW_fault_TB15}, \ref{fig:VIabc_AVG_fault_TB15}, and \ref{fig:Bfault_TB15}. The three-phase-to-ground fault occurs at 500~s and is removed after 0.15~s. The voltage drops to zero when a three-phase-to-ground fault occurs, as shown in Figs.~\ref{fig:VIabc_DSW_fault_TB15} and \ref{fig:VIabc_AVG_fault_TB15}. The terminal voltage dip is detected, and the grid-side controller's current limit functions begin restricting the injected active and reactive current during the fault. During this condition, the output power was reduced to zero, as shown in Fig.~\ref{fig:Bfault_TB15}. Because the wind turbine continues to absorb wind energy, the DC link power grows dramatically, resulting in a substantial increase in DC link voltage. In this scenario, the DC chopper is turned on to protect the DC link from damage caused by overvoltage, as shown in Fig.~\ref{fig:Bfault_TB15}(d). The DC link voltage and output power are restored to their nominal values after the fault is cleared. It is observed that the DSW model exhibits high-frequency harmonic distortion, and the AVG model can represent the same dynamic response as the DSW model. 
	\begin{figure}[!bt]
		\centering
		\includegraphics[width=0.78\linewidth]{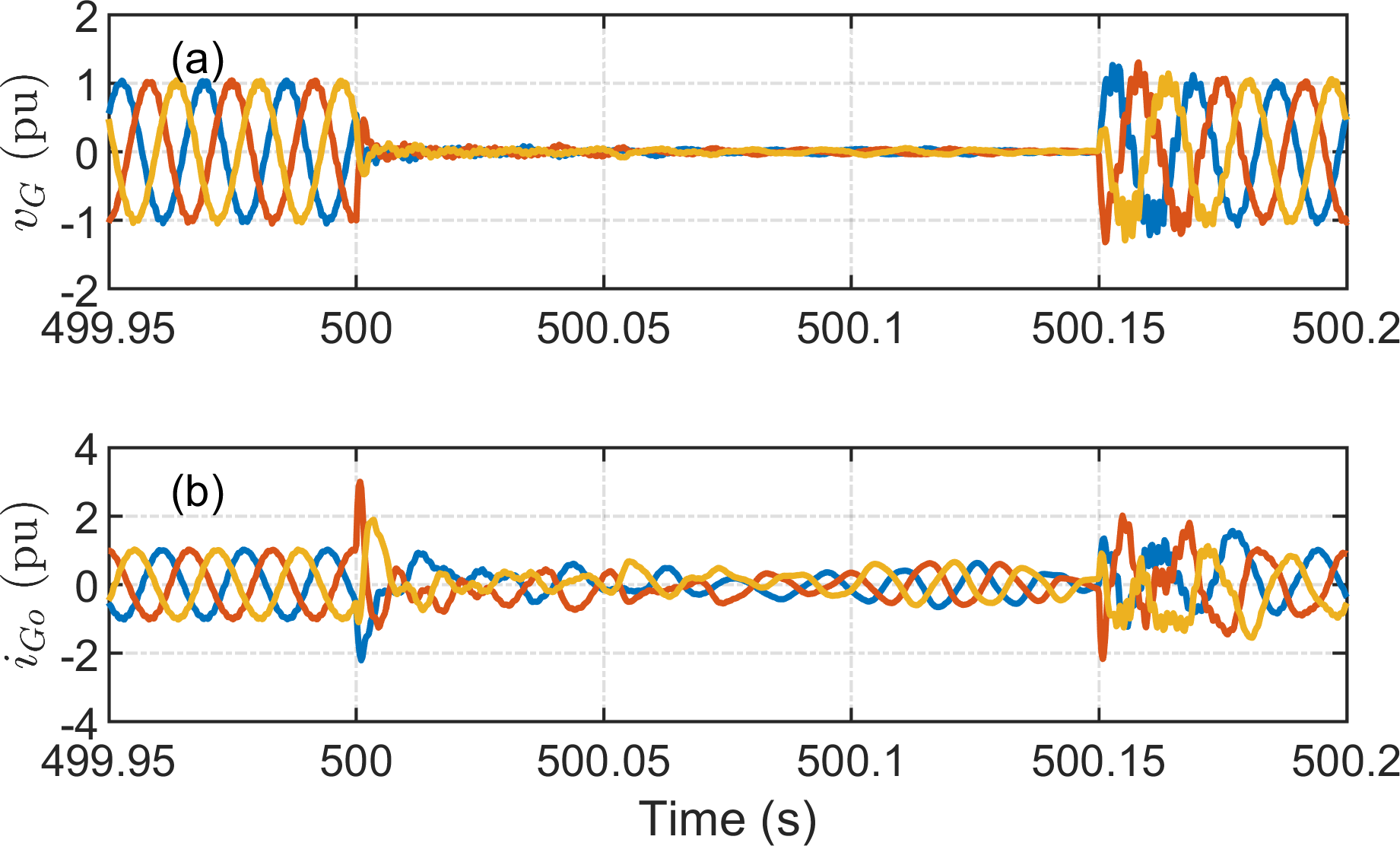}
		\caption{Transient three-phase voltage and current of DSW model measured at turbine terminal: (a) Voltage; (b) Current. Note: Blue line represents phase a, orange line represents phase b, and yellow line represents phase c.}
		\label{fig:VIabc_DSW_fault_TB15}
	\end{figure}
	
	\begin{figure}[!bt]
		\centering
		\includegraphics[width=0.78\linewidth]{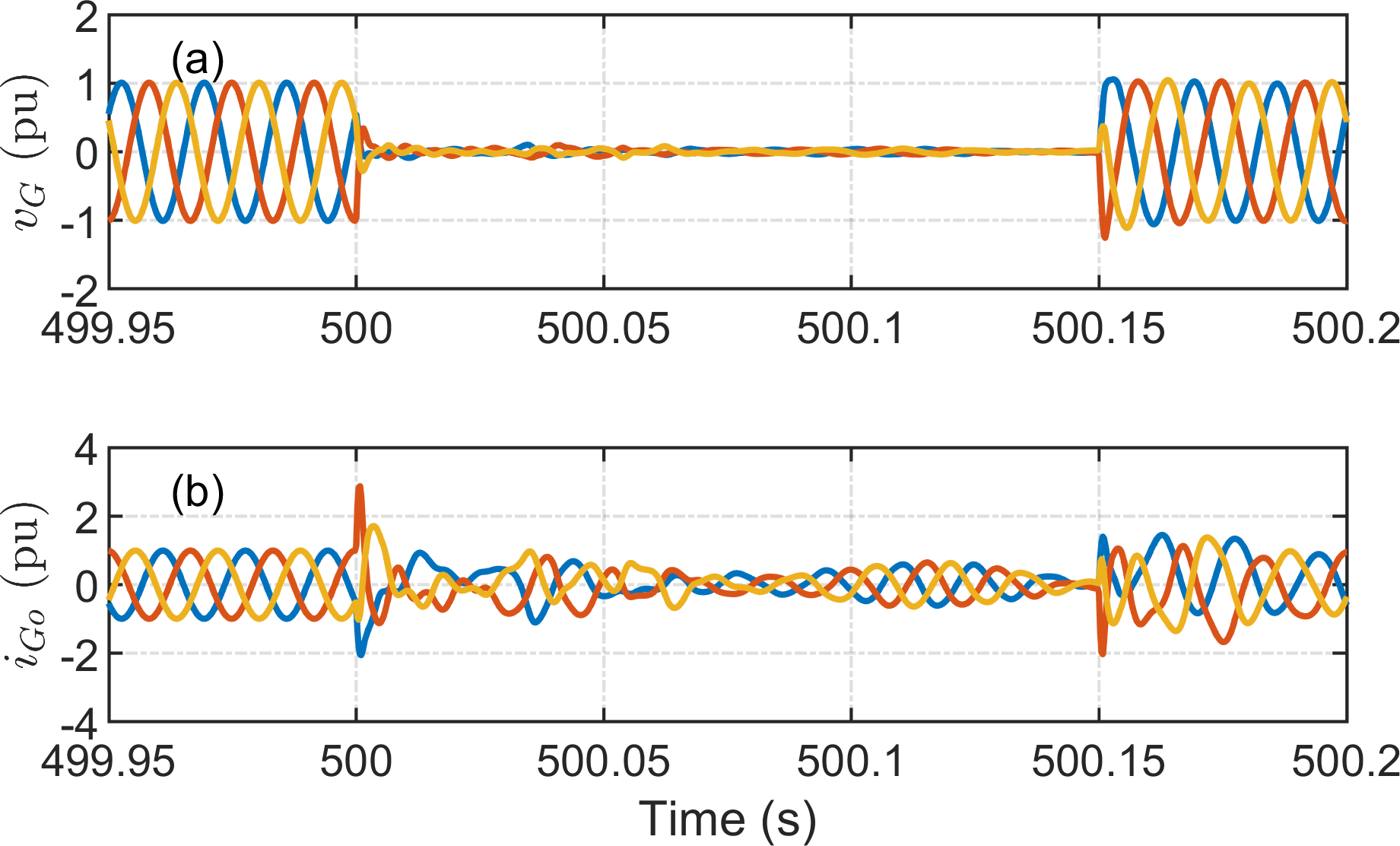}
		\caption{Transient three-phase voltage and current of AVG model measured at turbine terminal: (a) Voltage; (b) Current. Note: Blue line represents phase a, orange line represents phase b, and yellow line represents phase c.}
		\label{fig:VIabc_AVG_fault_TB15}
	\end{figure}
	
	\begin{figure}[!bt]
		\centering
		\includegraphics[width=0.78\linewidth]{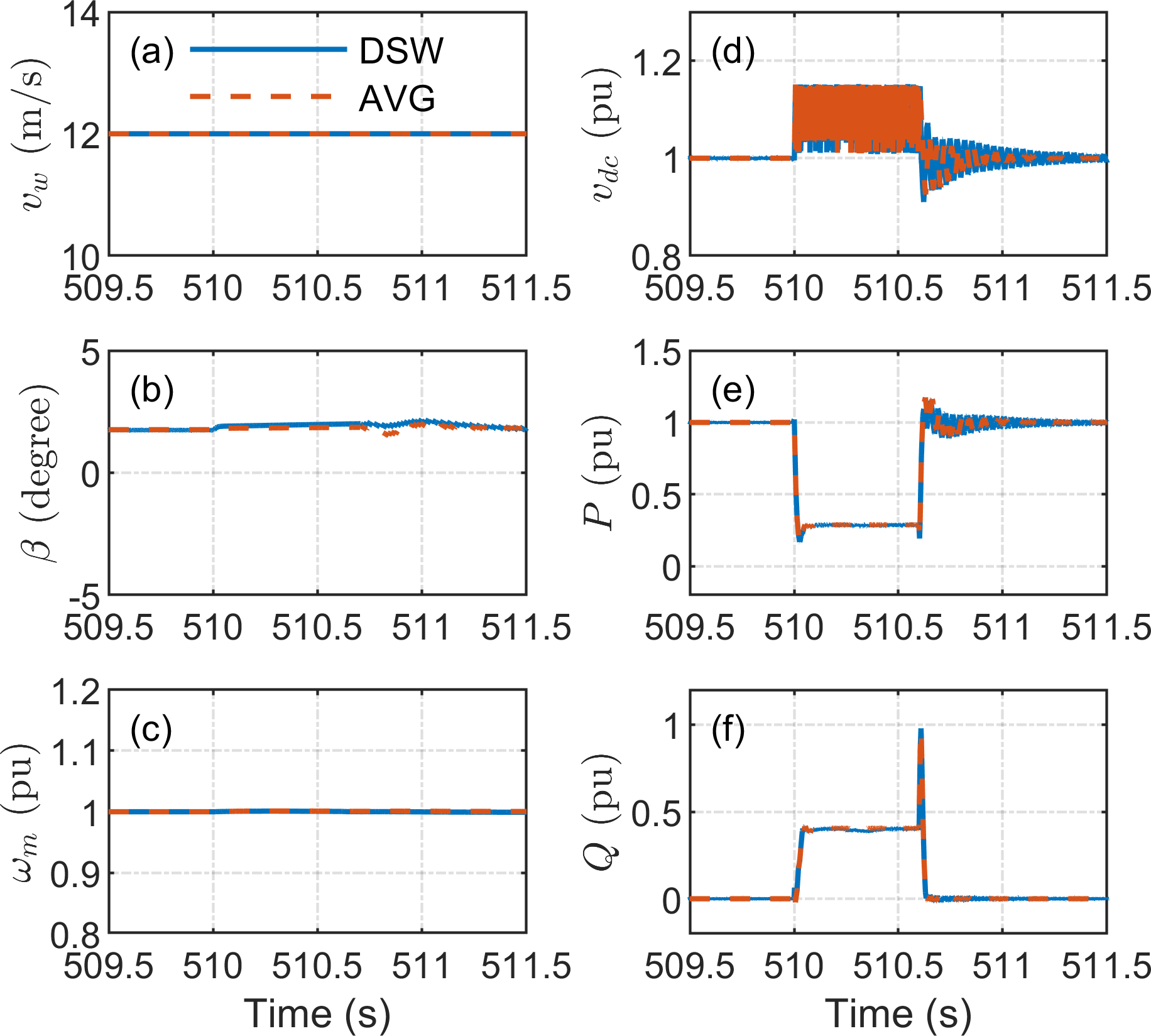}
		\caption{Balanced fault performance of 15MW model: (a) wind speed; (b) pitch angle; (c) rotor speed; (d) DC-link voltage; (e) output active power; (f) output reactive power.}
		\label{fig:Bfault_TB15}
	\end{figure}

	\subsection{Unbalanced Fault Studies}
	
	\begin{figure}[!bt]
		\centering
		\includegraphics[width=0.78\linewidth]{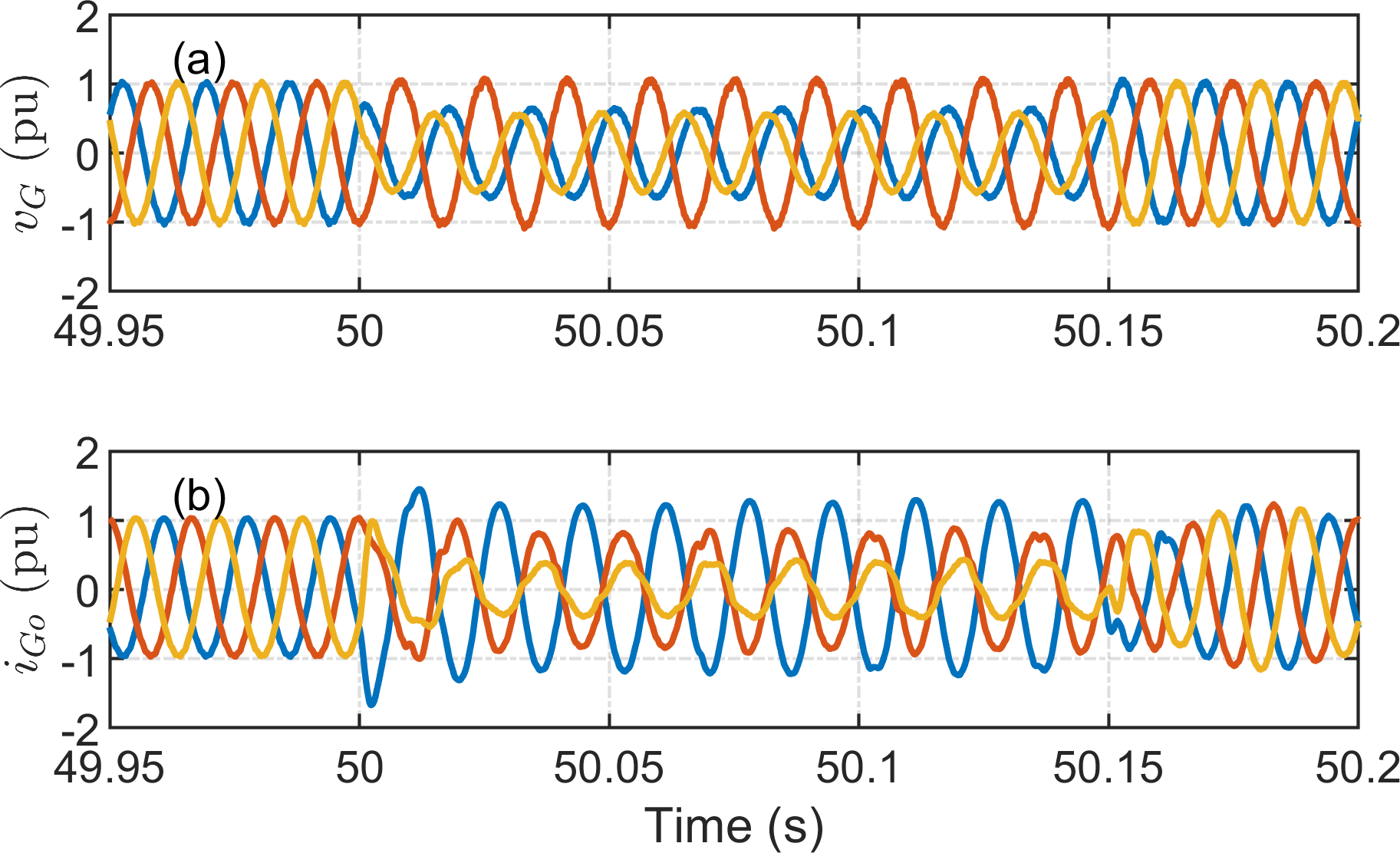}
		\caption{Transient voltage and current of typical control model under unbalanced fault: (a) Voltage; (b) Current. Note: Blue line represents phase a, orange line represents phase b, and yellow line represents phase c.}
		\label{fig:VIabc_Typ_Vdip0p35_TB15}
	\end{figure}
	
	\begin{figure}[!bt]
		\centering
		\includegraphics[width=0.78\linewidth]{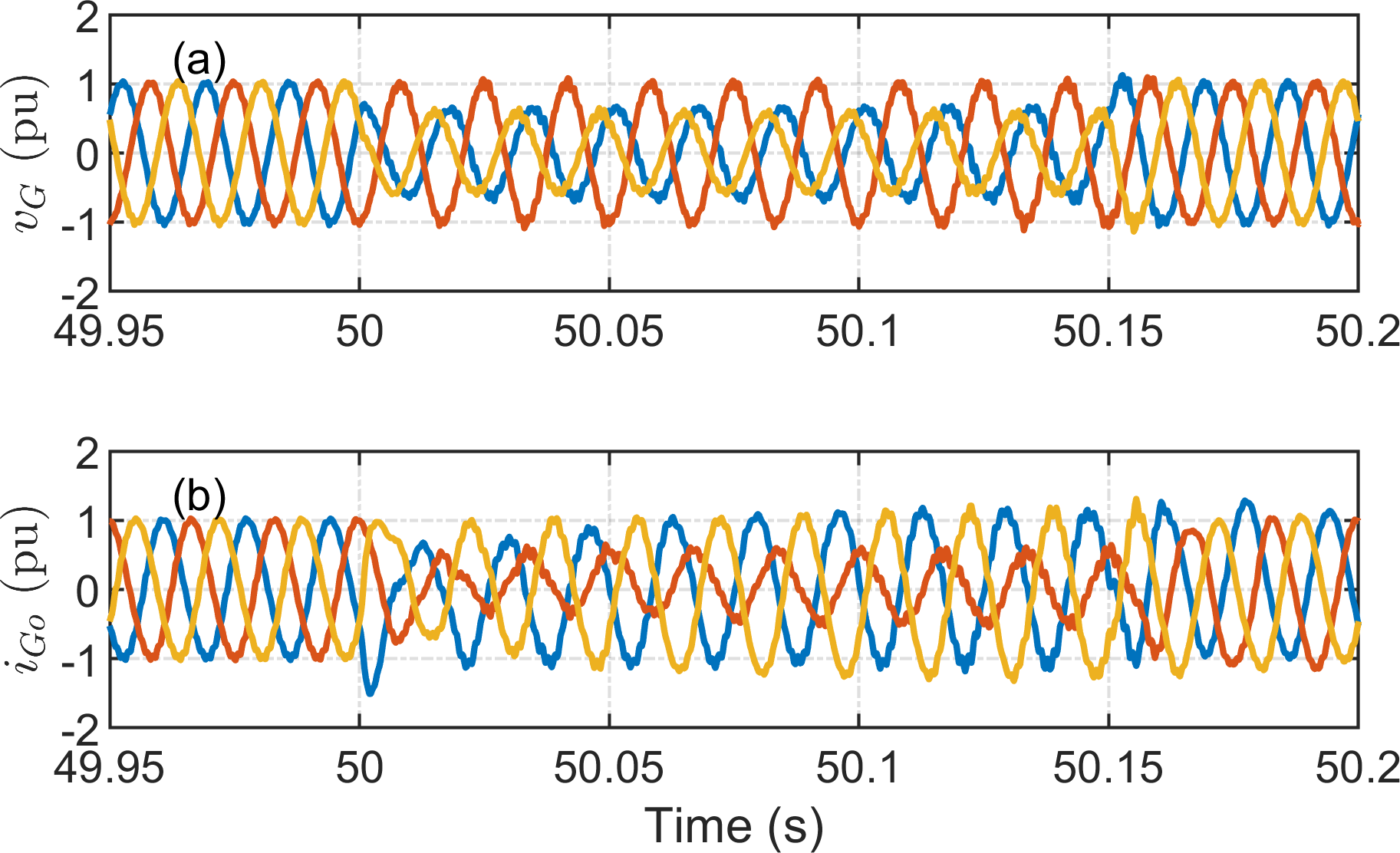}
		\caption{Transient voltage and current of proposed sequence-control model under unbalanced fault: (a) Voltage; (b) Current. Note: Blue line represents phase a, orange line represents phase b, and yellow line represents phase c.}
		\label{fig:VIabc_Seq_Vdip0p35_TB15}
	\end{figure}
	
	\begin{figure}[!bt]
		\centering
		\includegraphics[width=0.78\linewidth]{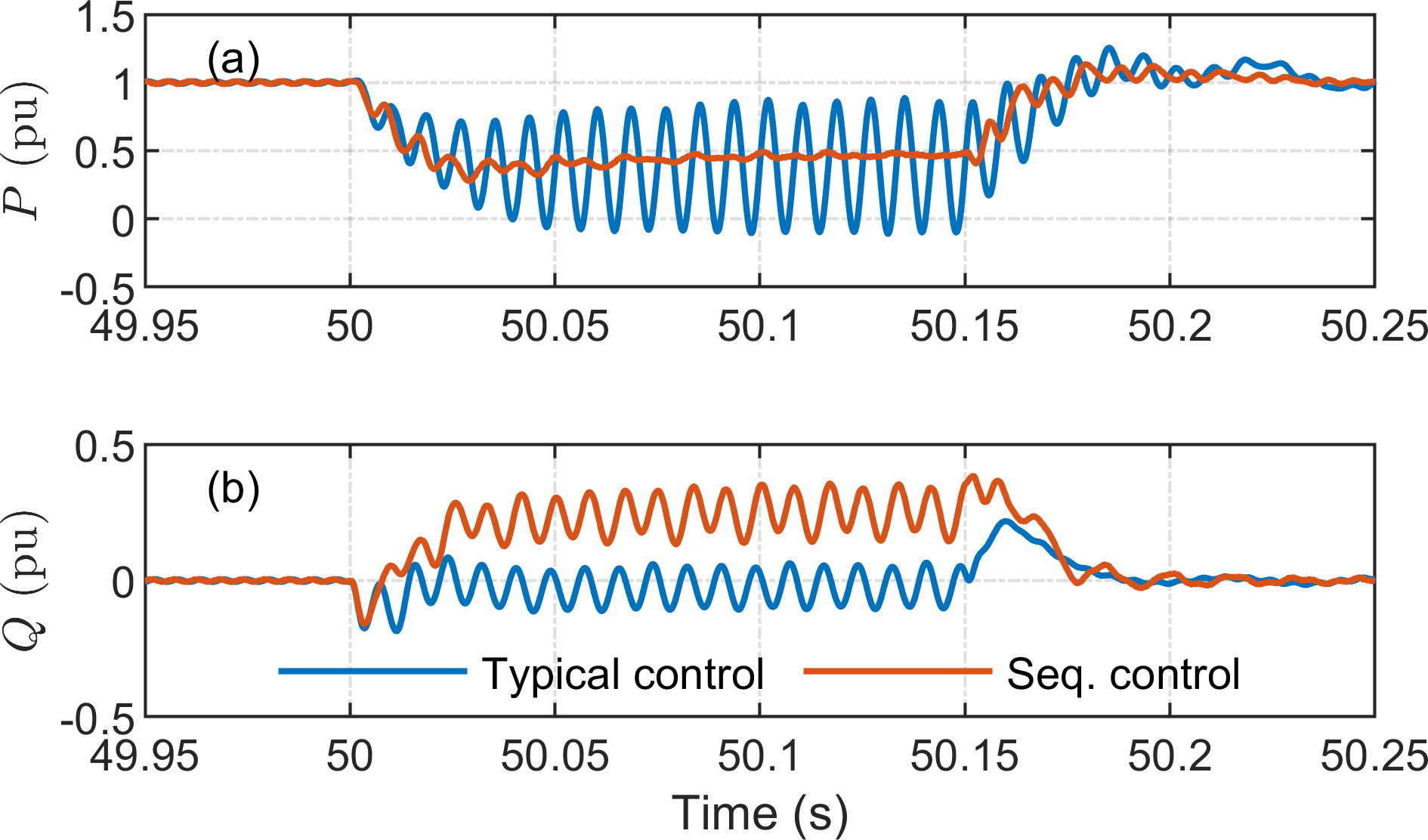}
		\caption{Comparison of transient power under unbalanced fault: (a) Active power; (b) Reactive power.}
		\label{fig:PQVdip0p35}
	\end{figure}
	
	
	The performance of the proposed turbine model is compared to that of a conventional turbine model without negative sequence control capabilities in unbalanced fault conditions. The phase-a-to-ground fault occurs on the high side of the step-up transformer from Fig.~\ref{fig:WTGMVsetup}. It should be noted that, while only the DSW models are compared in this section, the AVG models' performance is identical. The transient voltage and current measured at turbine terminals are shown in Figs.~\ref{fig:VIabc_Typ_Vdip0p35_TB15} and \ref{fig:VIabc_Seq_Vdip0p35_TB15}. The output current of the turbine under the proposed controller is different from the conventional control model because the proposed model can inject negative sequence current, as shown in Figs.~\ref{fig:VIabc_Typ_Vdip0p35_TB15}(b) and \ref{fig:VIabc_Seq_Vdip0p35_TB15}(b). The comparison of power during an unbalanced fault in Fig.~\ref{fig:PQVdip0p35} shows that the second-order harmonic oscillation in the active power component is damped out with the proposed sequence control. This is an advantage of the proposed turbine models. 
	
	\section{Conclusion}
	\label{sec:conclusion}
	This paper proposed real-time models of offshore wind turbines with the capabilities of limiting sequence currents and injecting negative sequence currents. The proposed models were developed in accordance with the standardized turbine models, making them appropriate for practical studies on interconnected wind farm systems. The turbine model validation studies, in which the proposed turbine models were validated against WECC generic turbine model, was presented to demonstrate the efficacy of the proposed models. The validation results reveal that the proposed models were aligned with the WECC generic turbine model. The real-time performance of both detailed switching and average turbine models was tested under normal and abnormal situations. \hlblue{The proposed average turbine model has the advantage of computational efficiency while precisely capturing the dynamic response of the detailed switching model, allowing for efficient real-time simulation of large-scale offshore wind farms.} The proposed models outperform the typical turbine models in damping second-order harmonic oscillations from active power due to their capacity to inject negative sequence currents under unbalanced conditions. The proposed models will be used to investigate large-scale wind farm systems in real time.

	\section*{Acknowledgment}

	This material is based upon research supported by, or in part by, the New York State Energy Research and Development Authority (NYSERDA) under award number 148516.

	\ifCLASSOPTIONcaptionsoff
	\newpage
	\fi

	\bibliographystyle{IEEEtran}
	\bibliography{References}

\begin{thebibliography}{10}
\providecommand{\url}[1]{#1}
\csname url@samestyle\endcsname
\providecommand{\newblock}{\relax}
\providecommand{\bibinfo}[2]{#2}
\providecommand{\BIBentrySTDinterwordspacing}{\spaceskip=0pt\relax}
\providecommand{\BIBentryALTinterwordstretchfactor}{4}
\providecommand{\BIBentryALTinterwordspacing}{\spaceskip=\fontdimen2\font plus
\BIBentryALTinterwordstretchfactor\fontdimen3\font minus
  \fontdimen4\font\relax}
\providecommand{\BIBforeignlanguage}[2]{{%
\expandafter\ifx\csname l@#1\endcsname\relax
\typeout{** WARNING: IEEEtran.bst: No hyphenation pattern has been}%
\typeout{** loaded for the language `#1'. Using the pattern for}%
\typeout{** the default language instead.}%
\else
\language=\csname l@#1\endcsname
\fi
#2}}
\providecommand{\BIBdecl}{\relax}
\BIBdecl

\bibitem{IRENA}
IRENA, ``Offshore innovation widens renewable energy options: Opportunities,
  challenges and the vital role of international co-operation to spur the
  global energy transformation,'' in \emph{(Brief to G7 policy makers),
  International Renewable Energy Agency, Abu Dhabi}, 2018.

\bibitem{gaertner2020definition}
E.~Gaertner, J.~Rinker, L.~Sethuraman, F.~Zahle, B.~Anderson, G.~Barter,
  N.~Abbas, F.~Meng, P.~Bortolotti, W.~Skrzypinski \emph{et~al.}, ``Definition
  of the iea 15-megawatt offshore reference wind turbine,'' 2020.

\bibitem{Vestas15MW}
\BIBentryALTinterwordspacing
C.~Richard. Vestas launches new 15mw offshore wind turbine with 236-metre
  rotor. [Online]. Available:
  \url{https://www.windpowermonthly.com/article/1706915/vestas-launches-new-15mw-offshore-wind-turbine-236-metre-rotor}
\BIBentrySTDinterwordspacing

\bibitem{8396277}
A.~S. {Trevisan}, A.~A. {El-Deib}, R.~{Gagnon}, J.~{Mahseredjian}, and
  M.~{Fecteau}, ``Field validated generic emt-type model of a full converter
  wind turbine based on a gearless externally excited synchronous generator,''
  \emph{IEEE Transactions on Power Delivery}, vol.~33, no.~5, pp. 2284--2293,
  2018.

\bibitem{5262959}
S.~M. {Muyeen}, R.~{Takahashi}, T.~{Murata}, and J.~{Tamura}, ``A variable
  speed wind turbine control strategy to meet wind farm grid code
  requirements,'' \emph{IEEE Transactions on Power Systems}, vol.~25, no.~1,
  pp. 331--340, 2010.

\bibitem{6520231}
K.~{Ma}, M.~{Liserre}, and F.~{Blaabjerg}, ``Power controllability of
  three-phase converter with unbalanced ac source,'' in \emph{2013
  Twenty-Eighth Annual IEEE Applied Power Electronics Conference and Exposition
  (APEC)}, 2013, pp. 342--350.

\bibitem{Karaagac2019EMT}
U.~{Karaagac}, J.~{Mahseredjian}, R.~{Gagnon}, H.~{Gras}, H.~{Saad}, L.~{Cai},
  I.~{Kocar}, A.~{Haddadi}, E.~{Farantatos}, S.~{Bu}, K.~W. {Chan}, and
  L.~{Wang}, ``A generic emt-type model for wind parks with permanent magnet
  synchronous generator full size converter wind turbines,'' \emph{IEEE Power
  and Energy Technology Systems Journal}, vol.~6, no.~3, pp. 131--141, 2019.

\bibitem{9067370}
R.~M. {Pindoriya}, B.~S. {Rajpurohit}, and A.~{Monti}, ``An investigative study
  of the pmsg based wind turbine using real time simulation,'' in \emph{2019
  8th International Conference on Power Systems (ICPS)}, 2019, pp. 1--6.

\bibitem{6869686}
S.~{Shah}, I.~{Vieto}, {Nian Heng}, and J.~{Sun}, ``Real-time simulation of
  wind turbine converter-grid systems,'' in \emph{2014 International Power
  Electronics Conference (IPEC-Hiroshima 2014 - ECCE ASIA)}, 2014, pp.
  843--849.

\bibitem{5342521}
W.~{Li}, G.~{Joos}, and J.~{Belanger}, ``Real-time simulation of a wind turbine
  generator coupled with a battery supercapacitor energy storage system,''
  \emph{IEEE Transactions on Industrial Electronics}, vol.~57, no.~4, pp.
  1137--1145, 2010.

\bibitem{5345723}
V.~{Jalili-Marandi}, L.~{Pak}, and V.~{Dinavahi}, ``Real-time simulation of
  grid-connected wind farms using physical aggregation,'' \emph{IEEE
  Transactions on Industrial Electronics}, vol.~57, no.~9, pp. 3010--3021,
  2010.

\bibitem{1022376}
{Yongsug Suh}, V.~{Tijeras}, and T.~A. {Lipo}, ``A control method in dq
  synchronous frame for pwm boost rectifier under generalized unbalanced
  operating conditions,'' in \emph{2002 IEEE 33rd Annual IEEE Power Electronics
  Specialists Conference. Proceedings (Cat. No.02CH37289)}, vol.~3, 2002, pp.
  1425--1430 vol.3.

\bibitem{8579187}
B.~{Mahamedi}, M.~{Eskandari}, J.~E. {Fletcher}, and J.~{Zhu}, ``Sequence-based
  control strategy with current limiting for the fault ride-through of
  inverter-interfaced distributed generators,'' \emph{IEEE Transactions on
  Sustainable Energy}, vol.~11, no.~1, pp. 165--174, 2020.

\bibitem{remtf2014wecc}
W.~REMTF, ``Wecc second generation of wind turbines models guidelines,''
  \emph{WECC, USA}, 2014.

\end{thebibliography}
	
\end{document}